\documentclass[11pt]{article}      
\usepackage{graphicx}
\author{Kris Krogh \\
Neuroscience Research Institute\\ University of California, Santa
Barbara, CA 93106, USA\\ email: k\_krogh@lifesci.ucsb.edu}

\title
{Iorio's ``high-precision measurement" of frame-dragging with the
Mars Global Surveyor}

\date{November 10, 2007}

\begin{document}

\maketitle

\begin{abstract}

\normalsize  In two analyses of orbital data from the Mars Global
Surveyor spacecraft, Iorio~\cite{li1,li2} has claimed confirmation
of the frame-dragging effect predicted by general relativity.
Initially to an accuracy of 6\%, and now 0.5\%, exceeding the
expected accuracy of NASA's Gravity Probe B.  It is shown his
results come from misinterpreting the MGS data and then altering a
key time period.
\end{abstract}

\vfill\eject

\section{Introduction}

The Mars Global Surveyor (MGS) reached Mars in September 1997. While
the main mission was imaging and mapping its surface, its magnetic
and gravitational fields were also mapped.  The latter began after
aerobraking in the Mars atmosphere put the probe in an approximately
circular polar orbit, at an altitude of 400 km, in February 1999.
The gravitational field data were obtained by observing its orbital
motion, via line-of-sight microwave Doppler and range measurements
between the probe and Earth stations.

As first described by Lens and Thirring~\cite{lt}, general
relativity predicts a rotating massive body would cause a rotational
dragging of nearby space-time. In two recent papers,
Iorio~\cite{li1,li2} argues this subtle effect can be detected in
the MGS orbit. Previous measurements have been reported by Ciufolini
and coauthors~\cite{ic,cp}, using the Earth-orbiting LAGEOS and
LAGEOS II satellites.  In that case, the Lens-Thirring effect tends
to be obscured by much larger ones, such as those of Earth and ocean
tides, and the accuracy of those results has been
questioned~\cite{ret,li3}.

A definitive finding is expected by year's end, with the
announcement of results from NASA's Gravity Probe B. That experiment
is an Earth-orbiting gyroscope, designed specifically to measure
Lense-Thirring precession, to an accuracy of 1\%. Iorio asserts he
has already achieved a measurement, first to 6\% and now 0.5\%
accuracy.  This has been well publicized.  (For example,
see~\cite{sc}.) Here we will review the methods and calculations
used to arrive at this extraordinary claim.

\section{Analysis of the MGS Orbital Data}

Iorio's Mars Global Surveyor papers~\cite{li1,li2} are based on
analysis of the MGS orbit by Konopliv \emph{et al.}~\cite{kysys}
Discussing their data, he writes in the first paper:

\begin{quote}
Here we are interested in particular in the out-of-plane portion of
its orbit. The  root-sum-square out-of-plane residuals over a
five-year time interval spanning from 10 February 2000 to 14 January
2005 are shown in figure 1. They have been determined in a
Mars-centered coordinate system. . . Due to improved modelling
(orientation, gravity, angular momentum wheel de-saturations, and
atmospheric drag), the average of such residuals amounts to 1.6 m.

This result can be well explained with the action of the
gravitomagnetic field of Mars on the orbit of MGS.
\end{quote}  The figure Iorio is referring to was reproduced from Konopliv \emph{et al.}  The original is shown as Figure 1 here.
(Unlike Iorio's version, it shows six years of data.)

To compare the average residual error in the MGS orbit with the
action of the gravitomagnetic field of Mars, Iorio starts with the
gravitomagnetic Lens-Thirring precession of a satellite orbit due a
rotating central body:
\begin{equation}
\dot\Omega_{\rm LT} \,=\: \frac{2\,G\,J}{c^2 a^3 (1 - e^2)^{3/2}}
\label{eq:1}
\end{equation}
(See Ciufolini and Wheeler~\cite{cw} for a derivation.)
$\dot\Omega_{\rm LT}$ refers to the secular rate of change in the
longitude of the orbit's ascending node, $G$ is the gravitational
constant, $J$ the rotating body's angular momentum, $c$ the speed of
light, $a$ and $e$ are the orbit's semi-major axis and eccentricity.

Referencing Christodoulidis {\em et al.}~\cite{dcc}, he finds that a
change in $\Omega_{\rm LT}$ can be expressed as change in the
position of the orbit's ascending node $N_{\rm LT}$ as
\begin{equation}
\Delta N_{\rm LT} \:=\: a\,\sqrt{1+\frac{\,e^2}{2}}\,\sin i
\,\Delta\Omega_{\rm LT}
\end{equation}
where $i$ is the orbital plane's inclination with respect to the
central body's equatorial plane. With $\Delta P$ representing a time
interval, he combines these expressions to get
\begin{equation}
\langle\Delta N_{\rm LT}\rangle \,=\: \frac{\,G\, J\, \Delta P \sin
i \sqrt{1 + \frac{\,e^2}{2}} }{c^2 a^2 ( 1-e^2)^{3/2}} \label{eq:3}
\end{equation}
The factor of 2 in Eq.~(\ref{eq:1}) is lost because $\langle\Delta
N_{\rm LT}\rangle$ represents the average change in position of the
ascending node over this time interval.

The result from this equation is compared in both papers to the 1.6
meter out-of-plane residual orbit error from Konopliv \emph{et al.}
The parameter values he uses are the same in both cases, except for
a $J$ which increases slightly, and the values for $\Delta P$, which
change from 5 years to 5 years and two months.  In his first paper,
he gets a predicted 1.5 meter average Thirring precession of the
ascending node, which he notes is within 6\% of the value from
Konopliv \emph{et al.}

Instead of meters, the result of the second paper is presented as a
ratio
\begin{equation}
\mu_{\rm meas}\equiv \left.\frac{\langle\Delta N_{\rm
res}\rangle}{\langle\Delta N_{\rm LT}\rangle}\right|_{\rm \Delta P}=
\,0.9937\pm 0.0053
\end{equation}
where the numerator refers to the 1.6 meter residual from Konopliv
\emph{et al.}  The predicted Lens-Thirring precession, in the
denominator, is 1.610 in this case.  Of the uncertainty on the
right, he attributes $\pm 0.0052$ to uncertainty in the Mars angular
momentum $J$, $\pm 0.0001$ to uncertainty in the constant $G$, and
\emph{none} to the residual $\langle\Delta N_{\rm res}\rangle$ from
observations of the MGS trajectory.

\begin{figure}
\begin{center}
\includegraphics[width=.78\textwidth]{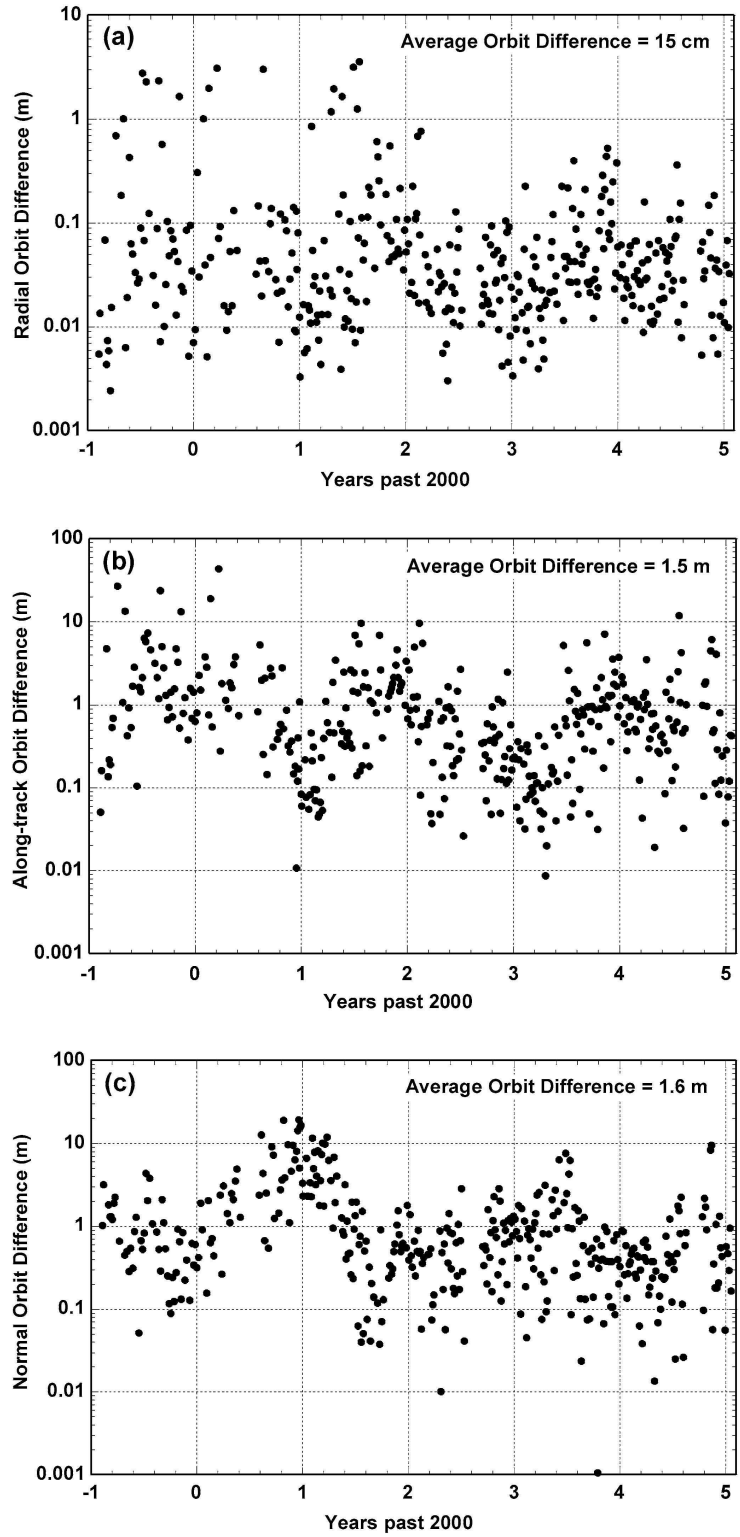}
 \caption{RMS orbit overlap differences from Konopliv \emph{et
al}~\cite{kysys}.  Reprinted with permission from Elsevier.}
\end{center}
 \label{fig:Konopliv}
\end{figure}

What information is actually shown in Figure 1 from Konopliv
\emph{et al}.? The number of data points is related to a spacecraft
trajectory model generated by the JPL Orbit Determination
Program~\cite{tdm}. While data from the entire mission is processed
to give a global best fit, the output is a sequence of separate
trajectory segments, ``data arcs," 4 to 6 (Earth) days in length.
Each plotted datum point corresponds to the intersection of two data
arcs.

Ideally, the modeled segments would fit together as a smooth,
continuous trajectory.  Since the modeling cannot be perfect, there
are mismatches. The orbital period of MGS is 2 hours, so a single
data arc consists of many orbits. As Konopliv \emph{et
al.}~\cite{kysys} describe, to gauge and minimize mismatches between
successive data arcs, each is extended to overlap the next by a full
orbit. The RMS difference in the spacecraft positions is then
computed over that orbit.  (Iorio refers to these as root-sum-square
residuals, although RMS indicates root-mean-square.)

These position differences are taken in three directions: radial,
along-track (the direction the spacecraft is moving), and normal to
the orbit plane.  Konopliv  \emph{et al.} plot these residual
modeling errors in a three-part figure.  Part (c), the ``normal
orbit difference," is the basis of Iorio's analysis, and is included
above. The original point made was that, as modeling of the Mars
gravity field and various perturbations have improved, the errors
have declined.  In this case, the normal average orbit error has
decreased from a previous 3 to 1.6 meters.

This does \emph{not} suggest a net precession of the orbit averaging
1.6 meters over the duration of the mission ($\Delta P$). It refers
to the average mismatch between successive data arcs. \emph{Over the
complete trajectory, the underlying directed errors might cancel, or
accumulate to a large number of meters.} Even individually, the
values do not specify orbital precessions.

The normal mismatch between overlapped orbits can be treated as
having two parts: an angular tilt of the orbit plane around some
axis through the Mars center of mass, plus a normal translation of
the orbit plane. (The latter effect would not be expected in a
rotationally symmetric system. For a planet with an irregular
gravitational field, it may be significant.) Alone, a 1 meter normal
translation of the orbit would give a 1 meter RMS difference, with
no angular precession of the orbit.

Assuming the normal orbit difference is only due to Lense-Thirring
precession, there would be no translation, and the corresponding
orbital tilt would be exactly around the Mars rotation axis. In this
case, the overlapped orbits would cross near the poles and separate
at the Mars equatorial plane. If the ascending node positions differ
there by 1 meter, the resulting RMS residual will be less, still not
corresponding to what is assumed in Iorio's calculation. Also, these
numbers have no sign to indicate a direction of precession.

Moreover, there is no reason to expect the value 1.6 is unique.
Lemoine \emph{et al.}~\cite{fgl} and \emph{Yaun et al.}~\cite{dny}
describe repeated decreases in the MGS orbit residuals as the
accuracy of Mars gravity maps and perturbation modeling have
improved. Nowhere in Konopliv \emph{et al.} is it suggested that the
present 1.6 m normal average orbit error could not be reduced by
further refinement of the existing orbit model.

Another crucial element of Iorio's analysis is his choice of the
time period $\Delta P$.  According to him, the out-of-plane orbit
error corresponds to a cumulative precession, increasing with time.
The 1.6 m error from Konopliv \emph{et al.} is based on 6 years of
data, shown in Figure 1. Without explanation, Iorio's first paper
takes the same error to apply to a 5-year period. (His figure omits
the data prior to 2000 to agree with that.)

In the second paper, he uses a $\Delta P$ equal to 5 years + 2
months, and obtains an order of magnitude better agreement between
general relativity and experiment. This is possible because, while
changing $\Delta P$ and with it the predicted error, he compares
that to the same 1.6 m measured error once again. This time he notes
\begin{quote}
By using a time span $\Delta P$ of slightly more than 5 years (14
November 1999$-$14 January 2005), so to remove the first months of
the mission more affected by orbital maneuvers and non-gravitational
perturbations, it is possible to obtain
\begin{equation}
\langle\Delta N_{\rm res}\rangle = 1.6\ {\rm m} \label{eq:5}
\end{equation}
Let us, now, compare the result of [Eq.~(\ref{eq:5})] with the mean
gravitomagnetic shift predicted by [Eq.~(\ref{eq:3})] over the same
time span. . .
\end{quote}
Two statistical tests are then used to show it is unlikely the close
correspondence between his predicted value and that observed is due
to chance.  (This conclusion seems justified.) Stressing the
objectivity of his findings, he also writes
\begin{quote}
Another important remark concerning our measurement is the following
one: the interpretation of the MGS residuals presented in this paper
and their production from the collected data were made by distinct
and independent subjects in different times; moreover, the authors
of~\cite{kysys} had not the Lense-Thirring effect in mind at all
when they performed their analysis. Thus, the possibility of any
sort of (un)conscious gimmicks, tricks or tweaking the data because
the expected answer was known in advance~\cite{sw} is a priori
excluded in this test. \end{quote} However, his results come from
misinterpreting the MGS data, and from his own values of $\Delta P$
-- twice altered in a way which gives closer agreement with general
relativity.

\section{Discussion}

It would be surprising if an accurate test of Lens-Thirring
precession (or any test of it) could be obtained from the Mars
Global Surveyor trajectory. Unlike some spacecraft, it was not
designed for measurement of its undisturbed inertial motion.  That
was possible, for example, with the Pioneer 10 and 11 space
probes~\cite{alllnt}. Those encountered no planetary atmospheres,
and were spin-stabilized, such that frequent use of thrusters was
not required.

In this case, AMD (angular momentum wheel desaturation) thrusters
were fired 4 or 5 times per day until September 2001, when that was
reduced to 1 or 2. Velocity uncertainties due to imprecision of the
thrust are partially corrected from observations of the spacecraft
motion, but that is limited by the one-dimensional nature of the
Doppler velocity measurements.

From Konopliv \emph{et al.}: ``Normal orbit error is the greatest,
and along-track is the least when the orbit plane is near edge-on as
viewed from Earth (e.g., January 2001). For this geometry, the
signature in the Doppler (which measures the velocity of the
spacecraft in the Earth-Mars direction) is greatest for motion in
the orbit plane, but is minimal for motion normal to the orbit
plane."

Referring to Figure 1, note the dips in the radial and along-track
differences centered around January 2001, and the corresponding
large hump in the normal orbit difference, where it increases by a
factor of ten. This obviously contributes to the 1.6 m average
normal orbit difference. We could agree with with Konopliv \emph{et
al.} that this is due to \emph{measurement} error. Certainly it
cannot be attributed to a larger general relativistic precession
occurring at that time.

Iorio's second paper states his value of $\Delta P$ was chosen ``so
to remove the first months of the mission more affected by orbital
maneuvers and non-gravitational perturbations." However, there was
no change in orbital maneuvers which would justify a time period
beginning in November 1999.  Also, Konopliv {\em et al.} point out
that noise in the Doppler velocity signal was {\em least} during the
first one and three quarters months, referred to as the ``Gravity
Calibration Orbit." This preceded deployment of the large high-gain
antenna, which increased atmospheric drag.

At the relatively low altitude of MGS, the Martian atmosphere is a
significant factor. From the spacecraft's asymmetric shape, drag due
to its forward motion would be expected to cause out-of-plane
forces. For a satellite in polar orbit, rotation of the upper
atmosphere would also tend to produce a force causing an orbital
precession. Solar heating of the spacecraft on emergence from the
Mars' shadow may be an additional factor.

None of the possible errors due to these or other orbital
perturbations are estimated by Iorio, or included in his
calculations. Following initial publication of this review on arXiv,
another by Sindoni, Paris and Ialongo~\cite{spi} has found the
uncertainty of the orbit's precession exceeds that claimed in
Iorio's second paper by at least a factor of $10000$. Also see
Felici~\cite{gf}.

\section{Conclusions}

Iorio makes the hypothesis that a certain measure of the error in
the modeled Mars Global Surveyor orbit is due to the Lense-Thirring
precession from general relativity.  He assumes this measure is
\emph{precisley} 1.6 meters.  This is shown to equal general
relativity's prediction, within a factor of 0.9937, and with an
uncertainty of $\pm 0.0053$.  However, close agreement was obtained
by calculating the predicted precession for a different time period
than the actual period of observations.

Also, where he assumes 1.6 meters represents the average of a
cumulative effect arising over 5 or 6 years, it actually
characterizes an average RMS orbital mismatch arising over two
``data arcs" -- with a total span in the range of 8 to 12 days. The
effects are generally not cumulative.  Finally, this quantity is not
a specific measure of movement of the orbit's ascending node, as
assumed in Iorio's calculations.

Other approaches to measuring the Lens-Thirring effect are
promising.  In addition to the pending result from Gravity Probe B,
a new experiment has been proposed by Turyshev \emph{et
al.}~\cite{sgt} This would observe gravitational bending of laser
light by the Sun, using two small, relatively inexpensive satellites
positioned on the far side of the Sun, and the International Space
Station.  Second-order bending would be measured for the first time,
as well as that due to frame dragging caused by the Sun's rotation.
The latter would be measured to 1\% accuracy.

\section*{Acknowledgments}

I am grateful to Stanley Robertson for first pointing out possible
effects of the Mars atmosphere on the MGS orbit, and for helpful
discussions. I thank Alexander Konopliv and E Myles Standish, for
confirming my description of the MGS orbit error residuals, and for
their support. Also the anonymous referees, for several useful
suggestions.

\vfill\eject

\end{document}